\documentclass[lettersize,journal]{IEEEtran}

\usepackage{amsmath,amsfonts}
\usepackage{algorithmic}
\usepackage{algorithm}
\usepackage{array}
\usepackage[caption=false,font=normalsize,labelfont=sf,textfont=sf]{subfig}
\usepackage{textcomp}
\usepackage{stfloats}
\usepackage{url}
\usepackage{verbatim}
\usepackage{graphicx}
\usepackage{cite}
\usepackage[switch]{lineno}
\usepackage{booktabs}
\usepackage{soul}
\usepackage{color} 

\begin{document}

%\linenumbers

\title{Parameter-efficient Dual-encoder Architecture with Differentiable Choquet Integral Fusion for Underwater Acoustic Classification}

\author{Amirmohammad~Mohammadi, Joshua~Peeples,~\IEEEmembership{Member,~IEEE}, and Alexandra~Van~Dine}%

\maketitle

%\thanks{Manuscript received Month Day, Year.}%

\begingroup
\def\thefootnote{\relax}

%\footnotetext{Manuscript received Month Day, Year.}

\footnotetext{DISTRIBUTION STATEMENT A. Approved for public release. Distribution is unlimited.
This material is based upon work supported by the Under Secretary of War for Research and Engineering under Air Force Contract No. FA8702-15-D-0001 or FA8702-25-D-B002. Any opinions, findings, conclusions or recommendations expressed in this material are those of the author(s) and do not necessarily reflect the views of the Under Secretary of War for Research and Engineering.

\copyright 2026 Massachusetts Institute of Technology.

Delivered to the U.S. Government with Unlimited Rights, as defined in DFARS Part 252.227-7013 or 7014 (Feb 2014). Notwithstanding any copyright notice, U.S. Government rights in this work are defined by DFARS 252.227-7013 or DFARS 252.227-7014 as detailed above. Use of this work other than as specifically authorized by the U.S. Government may violate any copyrights that exist in this work.}
\footnotetext{A. Mohammadi and J. Peeples are with the Department of Electrical and Computer Engineering, Texas A\&M University, College Station, Texas, 77843, USA, e-mails: (amir.m@tamu.edu and jpeeples@tamu.edu). A. Van Dine is with the Massachusetts Institute of Technology Lincoln Laboratory, Lexington, Massachusetts, 02421, USA, e-mail: (alexandra.vandine@ll.mit.edu).}
\endgroup

\begin{abstract}
Underwater acoustic classification has a wide array of oceanic applications, but faces challenges due to an increasingly complex acoustic environment. Waveform and spectrogram representations have been primarily used as acoustic data features for classification tasks in this domain. Spectrograms model harmonic dependencies, but these reduced representations can filter out acoustic features relevant for discrimination. While phase information from the waveform allows full characterization of the signal, the original waveform can be noisy and complex, rendering this representation difficult for models to process directly. This paper proposes a dual-encoder neural architecture to simultaneously process acoustic waveforms and spectrograms, leveraging pre-trained backbones and parameter-efficient fine-tuning modules, enabling a domain adaptation. To combine these adapted branches, a novel differentiable fuzzy aggregation mechanism based on the Choquet integral is introduced to balance the temporal and spectral representations. This fusion strategy not only yields higher classification accuracy but also provides interpretability. Specifically, by analyzing the learned fuzzy measures, insights are revealed about class-specific shifts in the network's representation reliance. By dynamically shifting attention to the representation least corrupted by potential asymmetric channel distortions, the proposed gating mechanism mitigates the non-stationary challenges of the underwater environment. Evaluations on the DeepShip and ShipsEar datasets demonstrate that the proposed architecture achieves classification improvements over independent single-encoder baselines, while simultaneously restricting the trainable parameter space. This mitigates the risk of overfitting on limited acoustic datasets while alleviating the computational costs associated with fully fine-tuning foundation models.
\end{abstract}


% --- MODULAR SECTIONS ---
\section{Introduction}

\IEEEPARstart{U}{nderwater} acoustic classification has a wide array of oceanic applications, ranging from autonomous exploration and seabed mapping to maritime security and environmental conservation \cite{testolin2020detecting, beckler2022multilabel}. Traditionally, these tasks have relied on manual frequency analysis and fixed mathematical models to identify predictable sound signatures based on the physical properties of the underwater environment \hbox{\cite{urick1983principles}}. Such examples include low frequency analysis and recording (LOFAR) spectra \hbox{\cite{luo2023survey}} and the Detection of Envelope Modulation on Noise (DEMON) \hbox{\cite{sichun2007demon, hashmi2023novel}.} However, conventional signal processing techniques, which often rely on hand-crafted features, face robustness difficulties under varying environmental distortions, necessitating frameworks capable of extracting invariant acoustic signatures \cite{jensen2011computational}.

Therefore, the paradigm in underwater acoustic classification has shifted from traditional heuristic-based methods toward deep learning architectures that automate feature extraction directly from data \cite{lecun2015deep}. Although deep learning has emerged as a transformative alternative to conventional signal processing by leveraging its representation learning capacity for autonomous feature extraction \cite{bianco2019machine, niu2022deep}, the deployment of reliable systems in real-world environments remains a challenge due to the stochastic nature of the underwater acoustic channel, which is characterized by multipath interference, frequency-dependent attenuation, and ambient noise \cite{xie2024adversarial, van2021deep}. In order to preserve high-resolution temporal dynamics and phase information which can be attenuated during the spectral transformation to a time-frequency data representation generation \cite{song2021deep}, two primary input representations have been used in this work. The first is a time-frequency representation that helps modeling spectral energy distributions and harmonic patterns which are correlated with narrowband machinery vibrations and engine tonals \cite{mckenna2012underwater, xu2024self}. The second is a direct representation of the acoustic waveform \cite{oord2016wavenet}. By operating in the temporal domain, the waveform representation preserves fine-grained phase dynamics, making it sensitive to broadband propeller cavitation and hydrodynamic flow events that might be smoothed over during spectral transformation \cite{purwins2019deep}.

While single-representation architectures have demonstrated success, relying on either temporal or spectral domains limits a model's ability to capture the complexity of ship-radiated noise \cite{yang2024underwater}. In response, recent research has explored fusion frameworks to integrate these complementary representations. However, using such frameworks, particularly those utilizing large-scale foundation models, raises questions on how to fuse the two information sources while limiting the additional computational cost. It has been shown that fully fine-tuning these foundation models on underwater datasets is computationally expensive and prone to overfitting due to limited data availability \cite{zhu2025ssast}. To mitigate these challenges, Parameter-Efficient Fine-Tuning (PEFT) methodologies, such as Low-Rank Adaptation (LoRA) and Histogram-based Parameter-efficient Tuning (HPT), have been introduced to enable efficient domain adaptation \cite{hu2021lora, mohammadi2025histogram}. By freezing the pre-trained backbones and optimizing only a minimal set of injected parameters, these methods reduce the computational footprint.

This work proposes a learnable, parameter-efficient, dual-encoder framework which leverages both raw waveforms and log-mel spectrograms. A differentiable Choquet integral layer is implemented at the decision level to learn the optimal weighting of each representation. The Choquet integral provides a way for quantifying the synergy and redundancy between information sources \cite{mccurley2023segmentation, grabisch1996linear}. This differentiable formulation facilitates the end-to-end optimization of these fusion weights, allowing the network to autonomously balance between input representations to optimize the objective function of the model.
In the work herein, evaluations on the underwater acoustic datasets demonstrate that this approach not only improves performance at limited parameter footprint but also provides interpretable insights into the relative importance of temporal and spectral features across different ship classes.
\section{Related Work}

Spectral-based deep learning models transform acoustic signals into two-dimensional representations, enabling the application of architectures such as the Audio Spectrogram Transformer (AST) and Residual Network (ResNet) Convolutional Neural Networks (CNNs) to identify spectral harmonics and patterns of relevance \cite{gong2021ast, xu2024self, hershey2017cnn}. On the other hand, waveform-based models utilize architectures such as wav2vec to process raw signals, preserving the fine-grained temporal structures and phase information \cite{hu2020auditory, baevski2020wav2vec}. While both paradigms have achieved success, single-representation approaches are inherently constrained by the available information, leading to performance degradation in high-noise environments common in datasets like ShipsEar and DeepShip \cite{santos2016shipsear, irfan2021deepship}.

The shift toward multi-representation fusion includes dual-branch networks extracting embeddings from both temporal and spectral domains, and aggregating via concatenation or additive fusion \cite{han2022underwater}. Studies have shown that the synergy between waveform dynamics and spectral energy allows for better discrimination between ship classes with overlapping harmonic profiles. However, traditional fusion methods remain static and lack the flexibility to adapt to changing signal-to-noise ratios (SNR) \cite{sun2023robust}. To address the inflexibility of static aggregation, recent methodologies have introduced dynamic weighting and multi-level interactions. For example, \cite{miao2025attentional} deploys an attentional multi-domain feature fusion network that adaptively selects and fuses time, frequency, and time-frequency representations using a three-branch attentional cross-correlation mechanism. Similarly, \cite{duan2025multilabel} establishes a multi-label recognition framework that utilizes deep equilibrium models and gated fusion mechanisms to iteratively update and enable low-to-high-level interactions between temporal and time-frequency representations.

Fuzzy measures, such as the Choquet integral, have been utilized to model non-linear interactions between overlapping features, and has shown promise in handling heterogeneous data in diverse fields like multi-model sentiment extraction \cite{mccurley2023segmentation, du2018multiple}. By leveraging the Choquet integral \cite{du2018multiple}, a given framework could theoretically favor spectral features when harmonic clarity is high, or shift to temporal features in the presence of broadband noise. Despite its mathematical advantages for handling heterogeneous data, the application of the Choquet integral for adaptive fusion in ship-radiated noise recognition remains unexplored. The subsequent section details the proposed framework, demonstrating how this fuzzy aggregation is adapted within a dual-encoder architecture.

\section{Method}

\begin{figure*}[ht]
\centering
\includegraphics[width=0.95\textwidth]{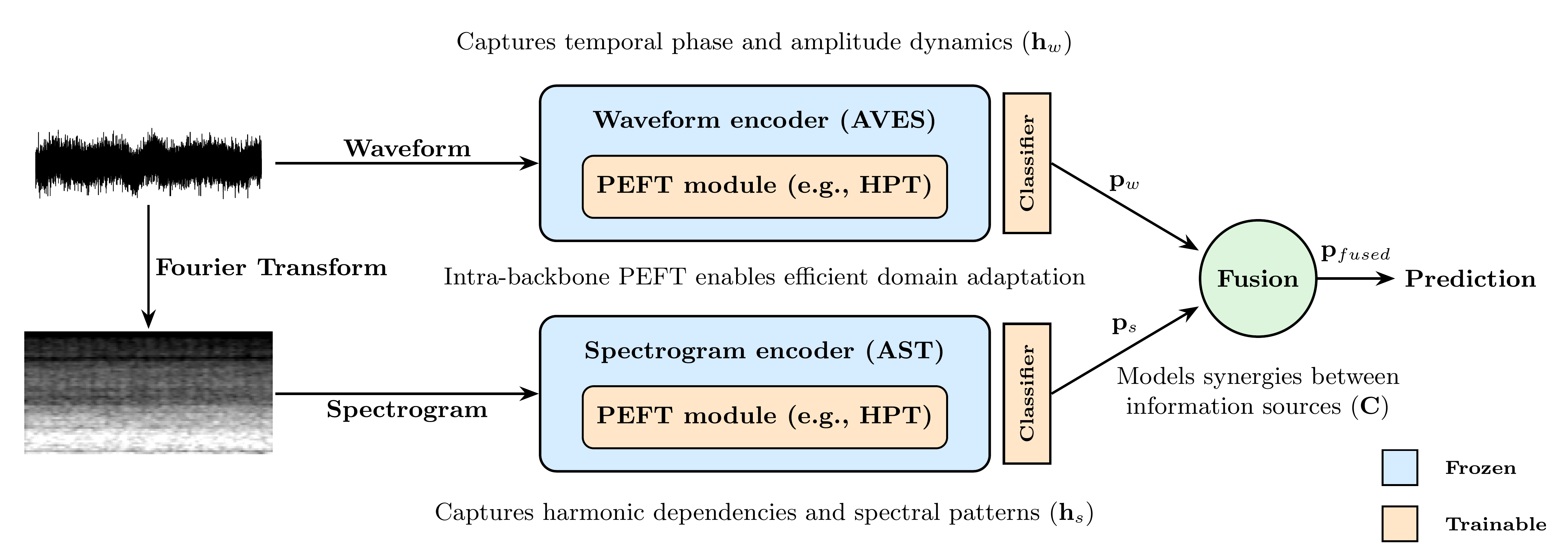}
\caption{The proposed dual-encoder architecture. The waveform and its corresponding spectrogram are processed by respective encoders. Parameter-efficient fine-tuning modules are integrated within the frozen backbones. The outputs are mapped by trainable classifiers and fused with the Choquet integral to yield the final prediction.}
\label{fig:method}
\end{figure*}

The proposed methodology, illustrated in Fig. \ref{fig:method}, leverages two parallel processing branches to simultaneously extract features from a one-dimensional acoustic waveform, denoted as $\mathbf{x}_w$, and its corresponding two-dimensional time-frequency spectrogram, denoted as $\mathbf{x}_s$. The temporal signal $\mathbf{x}_w$ is ingested by a waveform encoder to capture fine-grained phase and amplitude dynamics, resulting in a latent representation $\mathbf{h}_w$. Concurrently, the time-frequency spectrogram $\mathbf{x}_s$ is processed by a secondary encoder designed to model structural harmonic dependencies and spectral patterns, yielding the latent representation $\mathbf{h}_s$. These latent vectors are passed through branch-specific linear classification layers. The linear classifiers project the features into class logits, $\mathbf{z}_w$ and $\mathbf{z}_s$. To transform these logits into probability distributions for each feature, $\mathbf{p}_w$ (waveform) and $\mathbf{p}_s$ (spectrogram),  the standard softmax function is applied:

\begin{equation}
\mathbf{p}_w = \text{softmax}(\mathbf{z}_w),
\end{equation}
\begin{equation}
\mathbf{p}_s = \text{softmax}(\mathbf{z}_s).
\end{equation}

A primary contribution of this work is the introduction of a fully differentiable, decision-level fusion mechanism using the Choquet integral \cite{mccurley2023segmentation, grabisch1995fuzzy}. Standard fusion, such as the average, assumes additivity, meaning the importance of one feature does not depend on the presence of another. However, features can exhibit interactions: they may be redundant (overlapping information) or synergetic (more valuable together than apart). To model these interactions, fuzzy measures are used. A fuzzy measure $\mu$ assigns a weight not just to individual elements, but to every subset of inputs. The Choquet integral acts as the functional that aggregates inputs with respect to these non-additive measures. Mathematically, for a set of inputs $x = (x_{1}, x_{2}, \dots, x_{n})$, the Choquet integral is defined by first reordering the inputs such that $x_{(1)} \leq x_{(2)} \leq \dots \leq x_{(n)}$. The integral is then computed as:

\begin{equation}
C_{\mu}(x) = \sum_{i=1}^{n} (x_{(i)} - x_{(i-1)}) \cdot \mu(A_{(i)}),
\end{equation}

\noindent where $A_{(i)} = \{(i), \dots, (n)\}$ represents the set of indices corresponding to the largest remaining values, and $x_{(0)} = 0$. While this formulation allows for non-linear modeling of feature dependencies, it introduces a computational challenge for optimization as it requires the hard sorting of input values, a non-differentiable operation that disrupts gradient flow in deep neural networks. 

To resolve this limitation, a soft-sort gating mechanism based on the fuzzy measures $\mathbf{g}_w$ and $\mathbf{g}_s$ is proposed. These measures are learnable parameters that quantify the independent contribution and relative significance of each representation for each target class. These fuzzy measures are optimized on a per-class basis, allowing the network to establish tailored representation preferences for different acoustic signatures. To satisfy the mathematical constraints of fuzzy measures, they are bounded between zero and one via a sigmoid activation function. The proposed smooth soft-sort gate vector $\mathbf{s}$ is defined to conditionally route the predictions based on their relative magnitudes:
\begin{equation}
\mathbf{s} = \sigma(k(\mathbf{p}_w - \mathbf{p}_s)),
\end{equation}

\noindent where $\sigma$ denotes the sigmoid activation function and $k$ is a scaling hyperparameter that controls the steepness of the comparator, ensuring the gate approaches binary values for disparate probabilities while remaining differentiable. In the implementation, $k$ was set to $10.0$ to provide a sharp yet stable routing threshold during backpropagation. The fused distribution denoted as $\mathbf{C}$ represents the continuous Choquet integral formulation, and is computed element-wise across the class dimensions:

% \begin{equation}
% \mathbf{C} = \mathbf{s} \odot \left( \mathbf{p}_s + (\mathbf{p}_w - \mathbf{p}_s) \odot \mathbf{g}_w \right) + (1 - \mathbf{s}) \odot \left( \mathbf{p}_w + (\mathbf{p}_s - \mathbf{p}_w) \odot \mathbf{g}_s \right),
% \end{equation}

\begin{equation}
\begin{split}
\mathbf{C} &= \mathbf{s} \odot \left( \mathbf{p}_s + (\mathbf{p}_w - \mathbf{p}_s) \odot \mathbf{g}_w \right) \\
&\quad + (1 - \mathbf{s}) \odot \left( \mathbf{p}_w + (\mathbf{p}_s - \mathbf{p}_w) \odot \mathbf{g}_s \right),
\end{split}
\end{equation}

\noindent where $\odot$ represents the element-wise Hadamard product. When the waveform probability exceeds the spectrogram probability for a given class, $\mathbf{s}$ approaches one, and the fusion algorithm utilizes the waveform fuzzy measure, $\mathbf{g}_w$, to scale the probabilistic difference. Conversely, when the spectrogram is more confident, the system prioritizes $\mathbf{g}_s$. Since the mathematical aggregation $\mathbf{C}$ does not guarantee a sum of one, a final normalization step is applied. The final fused prediction vector $\mathbf{p}_{fused}$ is obtained by dividing $\mathbf{C}$ by the sum of its elements across all classes, ensuring a valid probability manifold.

To optimize the dual-encoder architecture, a composite loss function is formulated to simultaneously train the individual branches and the joint fusion mechanism. Let $\mathbf{y}$ represent the ground-truth class labels. The total multi-objective loss, $L_{total}$, is calculated as a weighted average of three separate cross-entropy terms:

\begin{equation}
L_{total} = \frac{1}{4} \Big( 2 \cdot \text{CE}(\mathbf{p}_{fused}, \mathbf{y}) + \text{CE}(\mathbf{p}_w, \mathbf{y}) + \text{CE}(\mathbf{p}_s, \mathbf{y}) \Big),
\end{equation}

\noindent where $\text{CE}$ denotes the standard cross-entropy function. By explicitly weighting the fused loss twice as heavily as the individual branch losses, the network is directed to prioritize the collaborative decision-making process while enforcing that each independent encoder learns discriminative representations. The weight of 2 assigned to the fused loss was empirically determined to yield optimal performance for the primary experiments. 

To perform domain adaptation without incurring expensive computational overhead, PEFT modules are integrated within each encoder. During the training phase, the pre-trained backbone parameters of both encoders remain frozen. Gradient descent optimization is thereby restricted to the injected tuning modules and the final linear classifiers. A shared-weight configuration is used for these tuning modules within each individual backbone such that the waveform encoder and the spectrogram encoder maintain their own distinct, independent sets of fine-tuning parameters, which are shared across the internal layers of their respective backbones. This design choice regularizes the adaptation process and constrains the overall trainable parameter space. The synergistic combination of this intra-backbone parameter-efficient adaptation, the differentiable Choquet decision fusion, and the multi-objective optimization forms a comprehensive framework for underwater acoustic classification. The subsequent section details the empirical experiments, including the datasets, processing configurations, and evaluations used to validate the approach.

\section{Experiment}

The proposed framework is evaluated on the DeepShip \cite{irfan2021deepship} and ShipsEar \cite{santos2016shipsear} datasets. DeepShip comprises four vessel classes: cargo, passenger ship, oil tanker, and tug. ShipsEar is organized into a five-class setting consisting of four vessel categories: small craft (fishing boats, trawlers, mussel boats, tugboats, and dredgers), small leisure and harbor craft (motorboats, pilot boats, and sailboats), passenger ferries (passengers), and large ocean-going ships (ocean liners and roll-on/roll-off vessels) as well as a background noise class (natural ambient noise). All audio recordings are downsampled to a unified sampling rate of 16 kHz and partitioned into non-overlapping segments of five seconds. To prevent data leakage, a strict recording-level random split of 70\%, 15\%, and 15\% is applied for the training, validation, and testing sets, respectively, with the exact split indices saved to ensure reproducible results. This configuration yields 23,362 training, 5,080 validation, and 5,067 testing samples for the DeepShip dataset, alongside 1,476 training, 268 validation, and 450 testing samples for the ShipsEar dataset. For the time-frequency branch, log-mel spectrograms are extracted directly from the segmented temporal signals. Prior to network ingestion, a per-sample z-score normalization is applied to both the acoustic waveforms and the log-mel spectrograms to stabilize the input distributions. 

During all experiments, the time-frequency branch utilizes the Audio Spectrogram Transformer (AST) \cite{gong2021ast} architecture initialized with weights pre-trained on AudioSet \cite{gemmeke2017audio} and ImageNet \cite{deng2009imagenet}. AST processes the input log-mel spectrogram as a sequence of discrete patches, leveraging multi-head self-attention mechanisms to capture long-range global context and structural harmonic dependencies. Concurrently, the waveform encoder is initialized using the Animal Vocalization Encoder based on Self-Supervision (AVES) architecture \cite{hagiwara2023aves}. The AVES architecture offers 4 pre-trained models determined by their specific training data composition. The AVES-core model is trained on FSD50K \cite{fonseca2021fsd50k}, a dataset comprising human and environmental sounds, combined with a core subset of AudioSet. Building upon this core, the AVES-bio and AVES-nonbio variants incorporate additional animal vocalizations and non-animal sounds, respectively, sourced from AudioSet and VGGSound \cite{chen2020vggsound}, while the AVES-all model integrates the full combination of these sources. For this study, the AVES-nonbio configuration is selected, since maritime ship acoustic signatures are mechanical and non-biological in nature, establishing an optimal initialization point for the subsequent fine-tuning on the underwater datasets.

Fully fine-tuned implementations of the AST and AVES architectures serve as state-of-the-art baselines. Benchmarking against these foundation models isolates the specific advantages of the proposed framework, which include significantly reducing the number of trainable parameters and enabling decision-level fusion interpretability, all while retaining competitive classification performance.
The classification performance of the independent single-encoder baselines and the proposed dual-encoder architecture is presented in Table \ref{tab:classification_results}. The LoRA adaptation was evaluated across query ranks ($r \in \{16, 32\}$), while the HPT was tested across histogram bin counts ($b \in \{16, 32, 64\}$). The empirical results demonstrate that integrating the waveform and spectrogram representations via the proposed decision-level fusion consistently yields higher test set accuracies across both the DeepShip and ShipsEar datasets compared to relying on either representation. Notably, in the linear probe setting, the dual-encoder structure compensates for the limited representation power of the fully frozen backbones. The dual-encoder PEFT modules close the performance gap with single-encoder full fine-tuning while constraining the trainable parameter space to a fraction of the full network capacity. This validates the efficacy of the framework to balance performance and computational efficiency.

\begin{table*}[t]
\caption{Accuracy (\%) for single and dual-encoder architectures evaluated on the DeepShip and ShipsEar datasets. Metrics are reported as the mean $\pm$ standard deviation across three experimental runs. The highest overall accuracy is denoted in \textbf{bold}, while the highest accuracy achieved specifically by a PEFT method is \underline{underlined} in each encoder type. In the parameter count column, the values separated by a slash (/) indicate the number of trainable parameters (in thousands) allocated for the DeepShip and ShipsEar datasets, respectively.}
\label{tab:classification_results}
\centering
\setlength{\tabcolsep}{8pt}
\begin{tabular}{llccc}
\toprule
\textbf{Encoder} & \textbf{Method} & \textbf{DeepShip} & \textbf{ShipsEar} & \textbf{\# Parameters (k)} \\
\midrule
Waveform     & Full Fine-tune & 67.9 $\pm$ 1.3 & 65.1 $\pm$ 1.9 & $90.2\times 10^{3}$ \\
             & Linear probe   & 65.6 $\pm$ 1.4 & 63.0 $\pm$ 0.6 & 3.1 / 3.8 \\
             & LoRA-q-16      & 66.0 $\pm$ 0.2 & 63.4 $\pm$ 1.1 & 27.7 / 28.4 \\
             & LoRA-q-32      & 65.2 $\pm$ 0.4 & 63.9 $\pm$ 0.8 & 52.2 / 53.0 \\
             & HPT-16         & 66.2 $\pm$ 1.2 & \underline{65.1 $\pm$ 0.2} & 15.4 / 16.2 \\
             & HPT-32         & \underline{66.7 $\pm$ 0.3} & 62.7 $\pm$ 0.7 & 27.7 / 28.5 \\
             & HPT-64         & 66.6 $\pm$ 0.4 & 64.3 $\pm$ 0.4 & 52.4 / 53.1 \\
\midrule
Spectrogram  & Full Fine-tune & 72.3 $\pm$ 1.4 & 67.3 $\pm$ 0.2 & $85.3\times 10^{3}$ \\
             & Linear probe   & 63.7 $\pm$ 0.7 & 60.5 $\pm$ 1.5 & 3.1 / 3.8 \\
             & LoRA-q-16      & 68.9 $\pm$ 1.0 & 66.1 $\pm$ 1.3 & 27.7 / 28.4 \\
             & LoRA-q-32      & 69.4 $\pm$ 1.2 & 67.0 $\pm$ 1.4 & 52.2 / 53.0 \\
             & HPT-16         & 69.3 $\pm$ 1.3 & 66.1 $\pm$ 1.3 & 15.4 / 16.2 \\
             & HPT-32         & 69.8 $\pm$ 0.7 & 66.6 $\pm$ 0.8 & 27.7 / 28.5 \\
             & HPT-64         & \underline{70.6 $\pm$ 1.8} & \underline{67.3 $\pm$ 1.2} & 52.4 / 53.1 \\
\midrule
Dual & Full Fine-tune & \textbf{74.7 $\pm$ 1.6} & \textbf{69.3 $\pm$ 0.6} & $175.4\times 10^{3}$ \\
             & Linear probe   & 67.7 $\pm$ 0.2 & 64.0 $\pm$ 0.8 & 6.2 / 7.7 \\
             & LoRA-q-16      & \underline{71.6 $\pm$ 1.0} & 68.3 $\pm$ 1.6 & 55.3 / 56.9 \\
             & LoRA-q-32      & 71.2 $\pm$ 1.2 & \underline{68.7 $\pm$ 0.9} & 104.5 / 106.0 \\
             & HPT-16         & 71.0 $\pm$ 0.8 & 65.9 $\pm$ 0.7 & 30.8 / 32.3 \\
             & HPT-32         & 71.1 $\pm$ 1.1 & 67.3 $\pm$ 1.2 & 55.4 / 57.0 \\
             & HPT-64         & 71.1 $\pm$ 0.4 & 68.1 $\pm$ 0.6 & 104.7 / 106.3 \\
\bottomrule
\end{tabular}
\end{table*}

To further contextualize these results, Table \hbox{\ref{tab:sota_classification_results}} benchmarks the dual-encoder architectures against a suite of fully fine-tuned, state-of-the-art single-encoder baselines. These include vision-based models (ResNet-50 \hbox{\cite{he2016deep}} and ConvNeXtV2-tiny \hbox{\cite{woo2023convnext}}), a pretrained audio neural network (CNN14 \hbox{\cite{kong2020panns}}), and a self-supervised audio spectrogram transformer (SSAST \hbox{\cite{gong2022ssast}}). The Dual Full Fine-Tune achieves the best average performance further supporting the use of both representations when compared to models that use either the waveform or spectrogram. However, the Dual Full Fine-Tune approach is the most computationally expensive model when observing the number of trainable parameters. When leveraging PEFT, the results demonstrate that while the Dual LoRA and Dual HPT frameworks optimize roughly 100k parameters, they outperform the majority of the fully fine-tuned single-encoder networks. This demonstrates the capability of the dual-encoder PEFT framework to deliver near state-of-the-art classification in underwater environments while maintaining computational efficiency.

\begin{table*}[ht]
\caption{Accuracy (\%) comparison of state-of-the-art single-encoder baselines against the proposed dual-encoder architecture on the DeepShip and ShipsEar datasets. Single-encoder models are fully fine-tuned. The highest overall accuracy is denoted in \textbf{bold}, while the second highest is \underline{underlined}. In the parameter count column, values separated by a slash (/) indicate the number of trainable parameters (in thousands) allocated for the DeepShip and ShipsEar datasets, respectively.}
\label{tab:sota_classification_results}
\centering
\setlength{\tabcolsep}{8pt}
\begin{tabular}{lccc}
\toprule
\textbf{Method} & \textbf{DeepShip} & \textbf{ShipsEar} & \textbf{\# Parameters (k)} \\
\midrule
ResNet-50 \cite{he2016deep} & 64.3 $\pm$ 2.2 & 58.3 $\pm$ 3.7 & $23.5\times 10^{3}$ \\
ConvNeXtV2-tiny \cite{woo2023convnext} & 71.1 $\pm$ 0.3 & 56.7 $\pm$ 3.5 & $27.9\times 10^{3}$ \\
SSAST \cite{gong2022ssast}  & 71.0 $\pm$ 0.9 & 63.2 $\pm$ 0.6 & $85.3\times 10^{3}$ \\
AST \cite{gong2021ast}      & \underline{72.3 $\pm$ 1.4} & 67.3 $\pm$ 0.2 & $85.3\times 10^{3}$ \\
CNN14 \cite{kong2020panns}  & 71.2 $\pm$ 1.5 & 61.6 $\pm$ 1.6 & $79.7\times 10^{3}$ \\
AVES \cite{hagiwara2023aves}& 67.9 $\pm$ 1.3 & 65.1 $\pm$ 1.9 & $90.2\times 10^{3}$ \\
Dual Full Fine-Tune         & \textbf{75.3 $\pm$ 1.2} & \textbf{69.6 $\pm$ 0.5} & $175.5\times 10^{3}$ \\
Dual LoRA                   & 71.2 $\pm$ 1.2 & \underline{68.7 $\pm$ 0.9} & 104.5 / 106.0 \\
Dual HPT                    & 71.1 $\pm$ 0.4 & 68.1 $\pm$ 0.6 & 104.7 / 106.3 \\
\bottomrule
\end{tabular}
\end{table*}

To interpret the decision-making capability of the dual-encoder framework, the learned fuzzy measures from the decision-level fusion mechanism are analyzed. Table \ref{tab:fusion_side_by_side} presents the optimized fusion weights. These parameters quantify the network's reliance on the waveform and spectrogram representations for each specific ship class. Under full fine-tuning, the fusion weights remain at 0.5 (initialized value) for both branches, indicating that when the encoders have maximum optimization capacity, the fusion weights do not reflect the representation importance while the accuracy is maximized on these datasets. However, the routing behavior shifts when the parameter space is constrained. In PEFT and linear probing, the fusion mechanism assigns higher fuzzy measure weights to the spectrogram branch, demonstrating that the time-frequency features extracted by the AST architecture are more robust. Despite this general trend, the learned weights exhibit class-wise variance. This is particularly evident in the DeepShip linear probe configuration for the Tanker class, where the waveform weight (0.22) is higher than the spectrogram weight (0.18). This confirms that the proposed differentiable fusion successfully identifies and prioritizes the optimal representation for specific acoustic ship signatures.

To interpret the routing behavior, an explainable artificial intelligence methodology was applied. Gradients of the target class logits ($\mathbf{z}_w$ and $\mathbf{z}_s$) were computed with respect to their corresponding input representations ($\mathbf{x}_w$ and $\mathbf{x}_s$). These gradients were multiplied by the original input signals to isolate predictive features, a feature attribution technique known as Gradient $\times$ Input~\cite{ancona2018towards}, and subsequently modulated by the network's sample-specific soft-sort gate ($s$). To quantify the overall magnitude of feature importance, the absolute value of these attributions is taken and subsequently normalized to a standard $[0, 1]$ scale for consistent visual comparison. Formally, the attribution heatmaps, denoted as $\mathbf{A}_w$ and $\mathbf{A}_s$, are calculated as:
\begin{equation}
\mathbf{A}_w = s \cdot \text{Normalize}\left( \left| \mathbf{x}_w \odot \frac{\partial z_w}{\partial \mathbf{x}_w} \right| \right),
\end{equation}
\begin{equation}
\mathbf{A}_s = (1 - s) \cdot \text{Normalize}\left( \left| \mathbf{x}_s \odot \frac{\partial z_s}{\partial \mathbf{x}_s} \right| \right).
\end{equation}

\begin{figure}[ht]
\centering
\includegraphics[width=0.9\columnwidth]{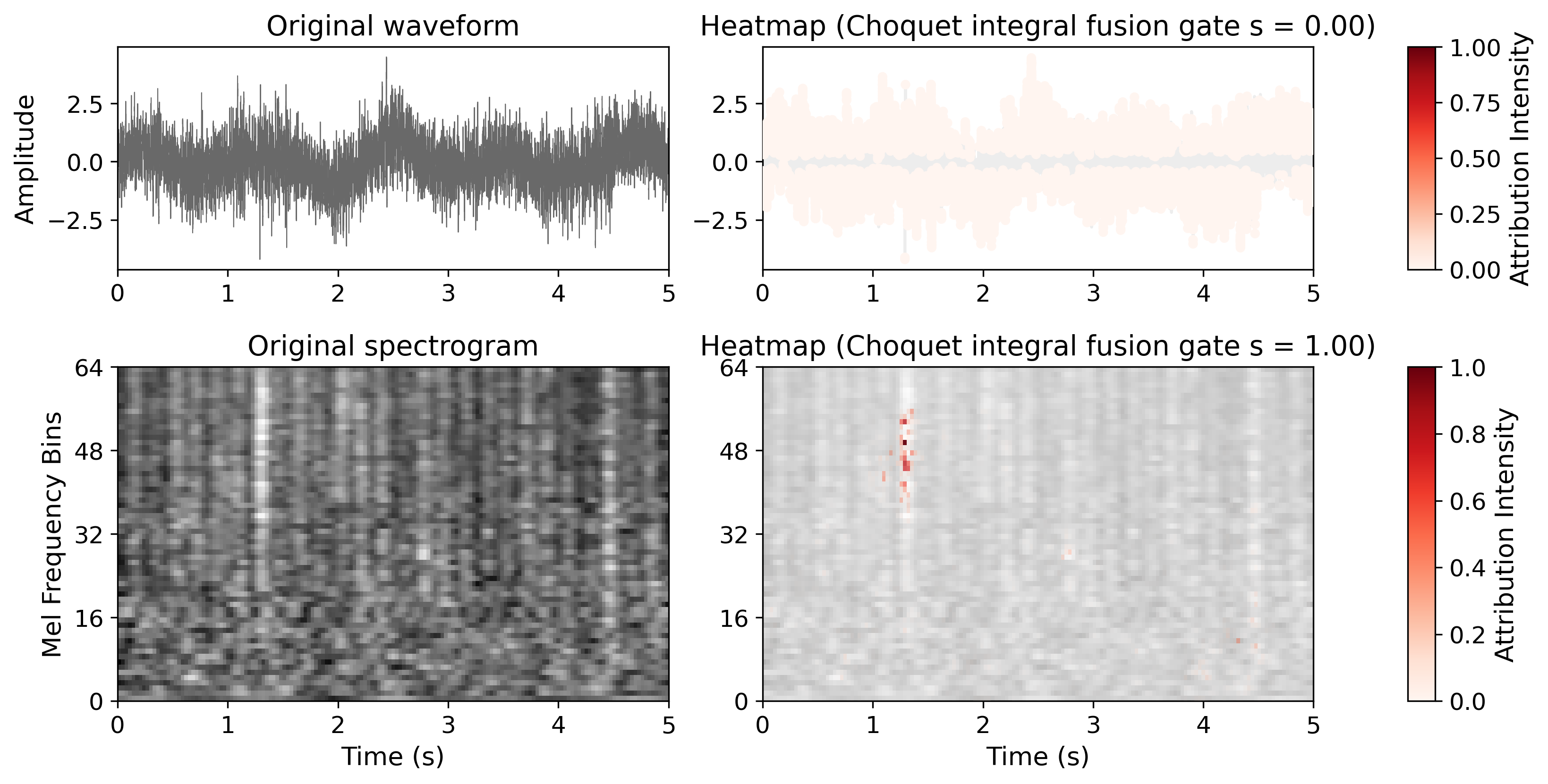}
\caption{Saliency map demonstrating dynamic representation routing for a Cargo sample, highlighting a spectral signature.}
\label{fig:xai_extreme_cargo}
\end{figure}

\begin{figure}[ht]
\centering
\includegraphics[width=0.9\columnwidth]{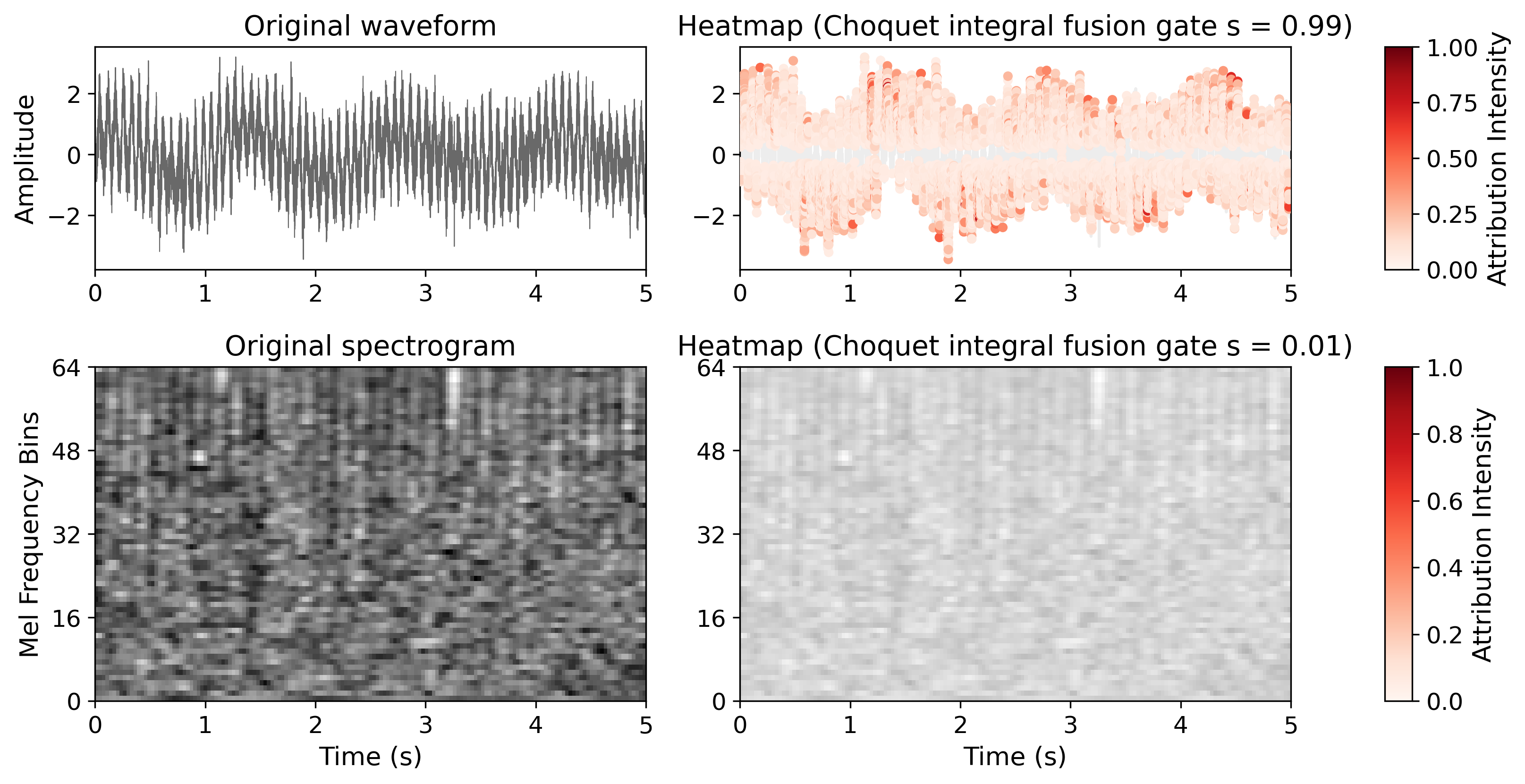}
\caption{Saliency map demonstrating dynamic representation routing for a Tanker sample, highlighting high-amplitude envelopes.}
\label{fig:xai_extreme_tanker}
\end{figure}

Figs.~\ref{fig:xai_extreme_cargo} and~\ref{fig:xai_extreme_tanker} illustrate two extreme routing scenarios under the linear probe configuration. For the Cargo sample (Fig.~\ref{fig:xai_extreme_cargo}, $s=0.00$), the decision-level fusion largely suppresses the temporal domain. Conversely, for the Tanker sample (Fig.~\ref{fig:xai_extreme_tanker}, $s=0.99$), the network shifts its reliance to the waveform. These visual explanations validate the dual-mode logic that the Choquet integral dynamically shifts attention to the most discriminative features in either the temporal or spectral representation. This dynamic routing capability is critical for robust deployment in the non-stationary underwater acoustic channel, where distortions such as multipath smearing of temporal waveforms and ambient noise masking of spectral tonals degrade acoustic features asymmetrically. Rather than statically fusing features, the proposed gating mechanism mitigates these distortions. By dynamically shifting reliance to the representation least corrupted by the local environment, the framework provides an operational advantage over single-encoder models.

\begin{table*}[t]
\caption{Learned Fuzzy Measure Weights Across Methods for DeepShip and ShipsEar Datasets}
\label{tab:fusion_side_by_side}
\centering
\setlength{\tabcolsep}{4pt} 
\begin{tabular}{l lcc c lcc}
\toprule
& \multicolumn{3}{c}{\textbf{DeepShip}} & & \multicolumn{3}{c}{\textbf{ShipsEar}} \\
\cmidrule{2-4} \cmidrule{6-8}
\textbf{Method} & \textbf{Class} & \textbf{Waveform} & \textbf{Spectrogram} & & \textbf{Class} & \textbf{Waveform} & \textbf{Spectrogram} \\
\midrule
Full Fine-tune & Cargo         & 0.50 $\pm$ 0.00 & 0.50 $\pm$ 0.00 & & A & 0.50 $\pm$ 0.00 & 0.50 $\pm$ 0.00 \\
               & Passengership & 0.50 $\pm$ 0.00 & 0.50 $\pm$ 0.00 & & B & 0.50 $\pm$ 0.00 & 0.50 $\pm$ 0.00 \\
               & Tanker        & 0.50 $\pm$ 0.00 & 0.50 $\pm$ 0.00 & & C & 0.50 $\pm$ 0.00 & 0.50 $\pm$ 0.00 \\
               & Tug           & 0.50 $\pm$ 0.00 & 0.50 $\pm$ 0.00 & & D & 0.50 $\pm$ 0.00 & 0.50 $\pm$ 0.00 \\
               &               &                 &                 & & E & 0.50 $\pm$ 0.00 & 0.50 $\pm$ 0.00 \\
\addlinespace
Linear probe   & Cargo         & 0.03 $\pm$ 0.01 & 0.53 $\pm$ 0.03 & & A & 0.42 $\pm$ 0.01 & 0.55 $\pm$ 0.01 \\
               & Passengership & 0.04 $\pm$ 0.01 & 0.46 $\pm$ 0.04 & & B & 0.44 $\pm$ 0.01 & 0.57 $\pm$ 0.00 \\
               & Tanker        & 0.22 $\pm$ 0.06 & 0.18 $\pm$ 0.01 & & C & 0.38 $\pm$ 0.01 & 0.53 $\pm$ 0.00 \\
               & Tug           & 0.04 $\pm$ 0.02 & 0.29 $\pm$ 0.04 & & D & 0.38 $\pm$ 0.00 & 0.52 $\pm$ 0.01 \\
               &               &                 &                 & & E & 0.40 $\pm$ 0.00 & 0.56 $\pm$ 0.01 \\
\addlinespace
LoRA           & Cargo         & 0.29 $\pm$ 0.02 & 0.64 $\pm$ 0.02 & & A & 0.43 $\pm$ 0.01 & 0.56 $\pm$ 0.01 \\
               & Passengership & 0.28 $\pm$ 0.01 & 0.65 $\pm$ 0.01 & & B & 0.43 $\pm$ 0.02 & 0.56 $\pm$ 0.01 \\
               & Tanker        & 0.41 $\pm$ 0.02 & 0.50 $\pm$ 0.01 & & C & 0.43 $\pm$ 0.01 & 0.56 $\pm$ 0.01 \\
               & Tug           & 0.30 $\pm$ 0.02 & 0.63 $\pm$ 0.02 & & D & 0.44 $\pm$ 0.01 & 0.55 $\pm$ 0.01 \\
               &               &                 &                 & & E & 0.46 $\pm$ 0.00 & 0.52 $\pm$ 0.00 \\
\addlinespace
HPT            & Cargo         & 0.15 $\pm$ 0.11 & 0.80 $\pm$ 0.11 & & A & 0.42 $\pm$ 0.01 & 0.58 $\pm$ 0.01 \\
               & Passengership & 0.15 $\pm$ 0.11 & 0.80 $\pm$ 0.12 & & B & 0.42 $\pm$ 0.01 & 0.59 $\pm$ 0.01 \\
               & Tanker        & 0.18 $\pm$ 0.13 & 0.75 $\pm$ 0.13 & & C & 0.37 $\pm$ 0.01 & 0.57 $\pm$ 0.01 \\
               & Tug           & 0.15 $\pm$ 0.11 & 0.79 $\pm$ 0.12 & & D & 0.37 $\pm$ 0.01 & 0.56 $\pm$ 0.01 \\
               &               &                 &                 & & E & 0.43 $\pm$ 0.01 & 0.54 $\pm$ 0.00 \\
\bottomrule
\end{tabular}
\end{table*}

\begin{figure}[ht]
\centering
\includegraphics[width=0.9\columnwidth]{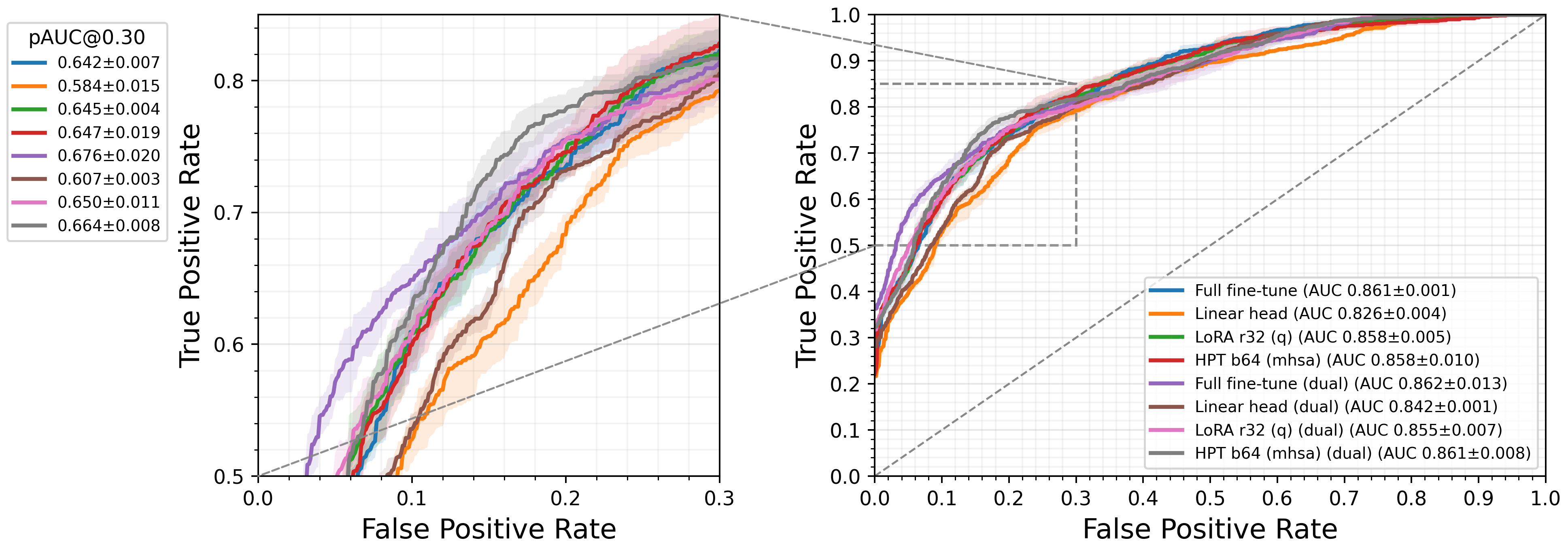}
\caption{Macro-averaged ROC curves and corresponding low-FPR context callout ($\text{pAUC}_{0.30}$) for the ShipsEar dataset, comparing single and dual-encoder configurations across various tuning strategies.}
\label{fig:roc_shipsear}
\end{figure}

\begin{figure}[ht]
\centering
\includegraphics[width=0.9\columnwidth]{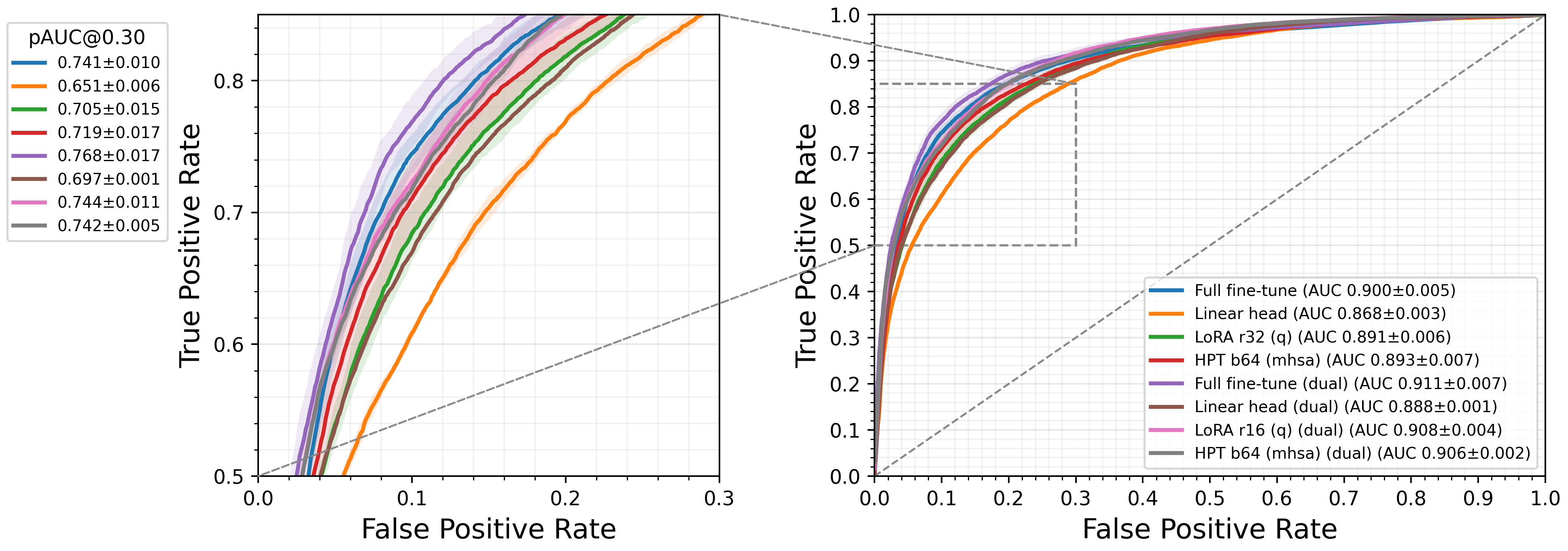}
\caption{Macro-averaged ROC curves and corresponding low-FPR context callout ($\text{pAUC}_{0.30}$) for the DeepShip dataset, comparing single and dual-encoder configurations across various tuning strategies}
\label{fig:roc_deepship}
\end{figure}

Figs. \ref{fig:roc_shipsear} and \ref{fig:roc_deepship} present the macro-averaged Receiver Operating Characteristic (ROC) curves for the ShipsEar and DeepShip datasets, respectively. These illustrate the True Positive Rate (TPR) against the False Positive Rate (FPR). To provide a granular view of model efficacy in low false-alarm operational regions, each figure includes a context callout panel zooming into the $0.00 \le \text{FPR} \le 0.30$ range. This allows for a detailed analysis of the partial Area Under the Curve ($\text{pAUC}_{0.30}$) alongside the standard global AUC metric. Across both datasets, the performance is improved when transitioning from the single-encoder to the dual-encoder representation, an effect most pronounced when utilizing the linear head. Furthermore, PEFT methods demonstrate statistically comparable performance to full fine-tuning. Within the ShipsEar evaluations, the dual-mode HPT achieves a $\text{pAUC}_{0.30}$ that is statistically significantly higher than that of the full fine-tuning method in single mode ($0.664 \pm 0.008$ vs. $0.642 \pm 0.007$), demonstrating the superior representational capacity of the dual-encoder architecture when constrained to low-false-positive acoustic scenarios.

To evaluate the representational similarity between the PEFT methods and the fully fine-tuned baseline across network depths, Centered Kernel Alignment (CKA) is employed \cite{kornblith2019similarity}. The CKA metrics are computed by extracting and mean-pooling the hidden states from each encoder layer. For two centered feature matrices, $X_c$ and $Y_c$, representing the layer activations of two models, the linear CKA is calculated as:
\begin{equation}
\text{CKA}(X, Y) = \frac{\|X_c^T Y_c\|_F^2}{\|X_c^T X_c\|_F \|Y_c^T Y_c\|_F},
\end{equation}
\noindent where $\|\cdot\|_F$ denotes the Frobenius norm. This formulation quantifies how closely the intermediate representations of the PEFT models align with a fully fine-tuned baseline.

\begin{figure}[ht]
\centering
\includegraphics[width=0.8\columnwidth]{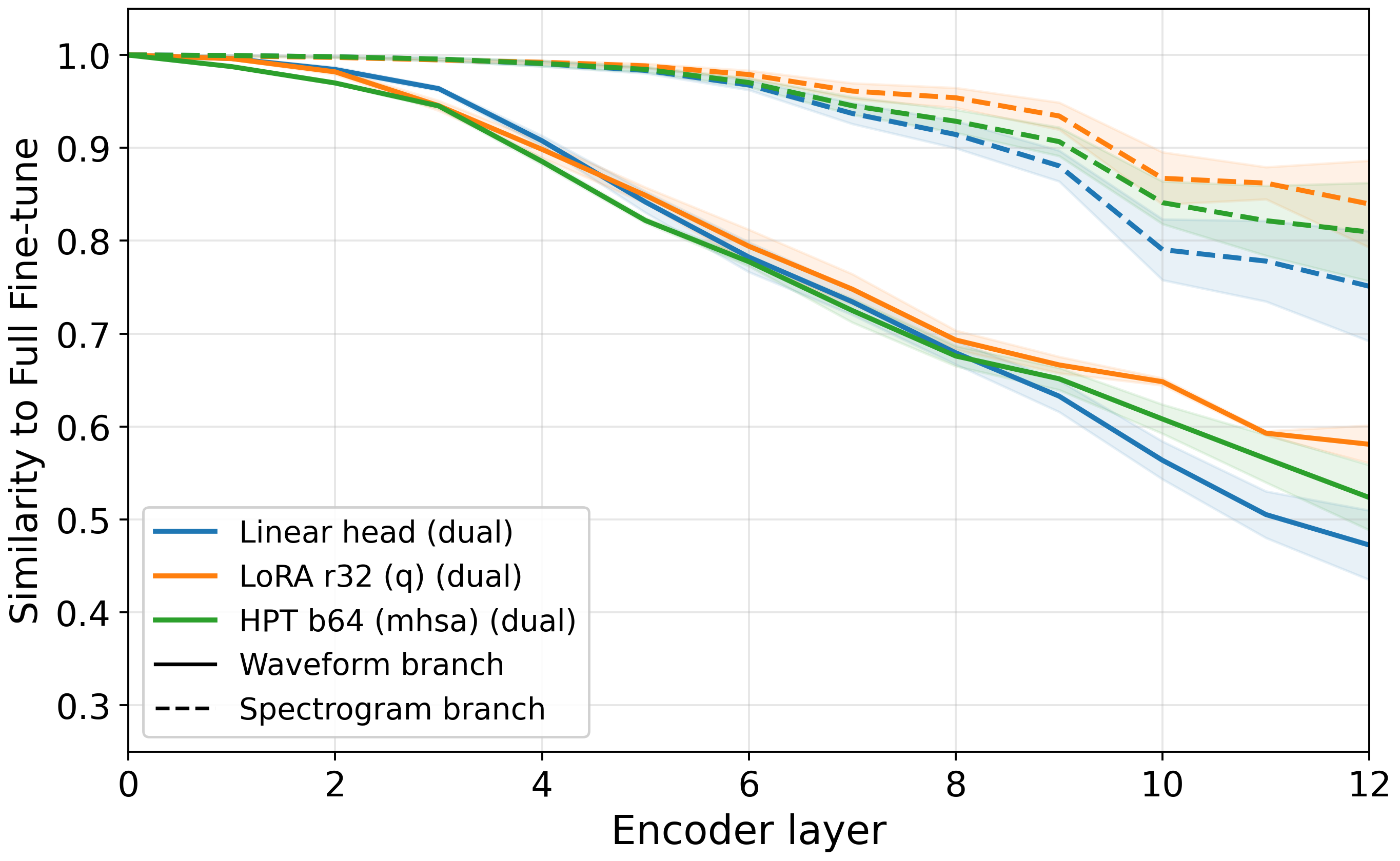}
\caption{Layer-wise Linear CKA similarity for the ShipsEar dataset. The figure compares the representational alignment of various methods against a fully fine-tuned baseline across the encoder layers.}
\label{fig:cka_shipsear}
\end{figure}

\begin{figure}[!ht]
\centering
\includegraphics[width=0.8\columnwidth]{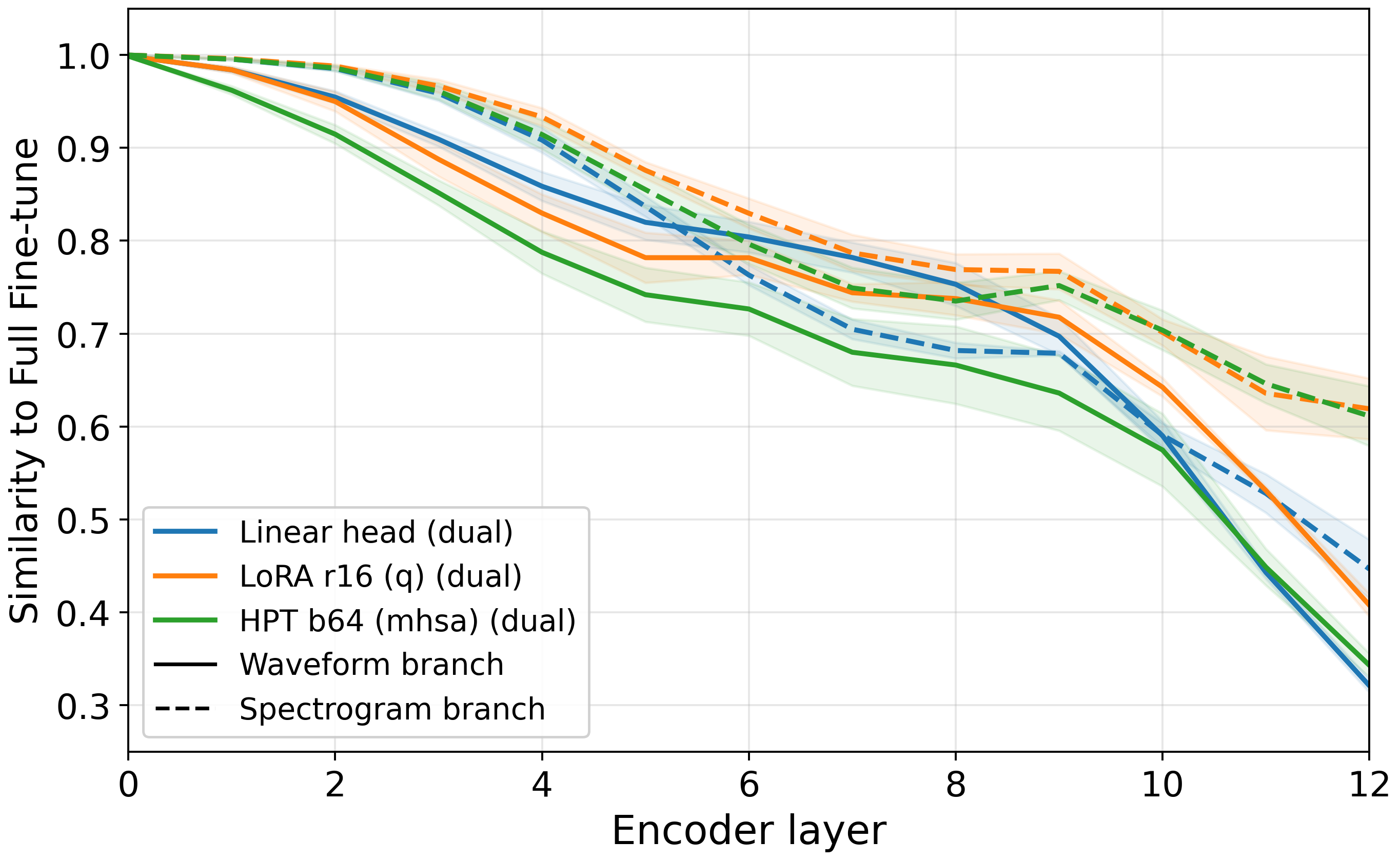}
\caption{Layer-wise Linear CKA similarity for the DeepShip dataset. The plot illustrates the divergence of representations from the fully fine-tuned baseline.}
\label{fig:cka_deepship}
\end{figure}

Figs. \ref{fig:cka_shipsear} and \ref{fig:cka_deepship} present the layer-wise CKA similarity for the ShipsEar and DeepShip datasets, respectively. A decreasing trend is observed as the network depth increases. This phenomenon indicates that early layers, which typically capture low-level features, remain relatively stable under constraints. Conversely, deeper layers, generally responsible for task-specific abstractions, diverge more from the fully fine-tuned baseline. Furthermore, the waveform branch exhibits lower CKA similarity than the spectrogram branch. This suggests that adapting representations directly from waveforms is more challenging. Among the evaluated adaptation strategies, the linear head demonstrates the lowest similarity, followed by HPT, while LoRA achieves the highest alignment to full fine-tuning. These results suggest that representational drift is concentrated in deeper layers, indicating that future adaptation techniques should utilize asymmetric parameter allocation by focusing tuning budgets on the final layers. Furthermore, the lower CKA in the waveform branch identifies it as a primary adaptation bottleneck, implying that it requires higher-capacity tuning (e.g., higher LoRA ranks) compared to the spectrogram branch to achieve optimal cross-representation alignment in maritime environments.

\section{Conclusion}

This paper presented a novel parameter-efficient dual-encoder framework for underwater acoustic classification. By simultaneously processing raw one-dimensional acoustic waveforms and two-dimensional time-frequency spectrograms, the proposed system effectively leverages both temporal dynamics and spectral dependencies. A primary contribution of this work is the introduction of a fully differentiable decision-level fusion mechanism based on the Choquet integral. Through the implementation of a smooth soft-sort gating mechanism, the architecture facilitates the end-to-end optimization of fuzzy measures. This enables the network to dynamically route predictions based on representation-specific confidence for individual ship classes. Furthermore, the integration of intra-backbone shared-weight PEFT modules ensures that the dual-encoder system adapts to specialized underwater domains without incurring the prohibitive computational footprints associated with full network fine-tuning.

Comprehensive empirical evaluations conducted on the DeepShip and ShipsEar datasets validated the efficacy of the proposed methodology. The dual-encoder architecture consistently outperformed independent single-encoder baselines, confirming the synergistic advantages of joint temporal and spectral acoustic feature capture. Notably, the application of PEFT techniques achieved highly competitive classification accuracies while reducing the trainable parameter space to a fraction of the baseline capacity. Finally, an analysis of the learned Choquet fuzzy measures provided valuable interpretability to the algorithmic decision-making process, revealing a dynamic shift in representation reliance when network capacity is constrained. These findings establish a computationally efficient framework for advanced maritime acoustic recognition. Future work will perform comprehensive evaluations to assess the computational cost and hardware efficiency of the framework. Following these evaluations, subsequent studies will investigate the application of quantization methods to the PEFT-integrated frozen backbones to minimize memory overhead. 

\section*{Acknowledgments}
Portions of this research were conducted with the advanced computing resources provided by Texas A\&M High Performance Research Computing.

% All figures are in the Figures folder.

% --- REFERENCES SECTION ---

% endfloat/captionsoff
\ifCLASSOPTIONcaptionsoff
  \newpage
\fi

\newpage
\bibliographystyle{IEEEtran}
\bibliography{ref}

% --- BIOGRAPHIES ---

\begin{IEEEbiography}[{\includegraphics[width=1in,height=1.25in,clip,keepaspectratio]{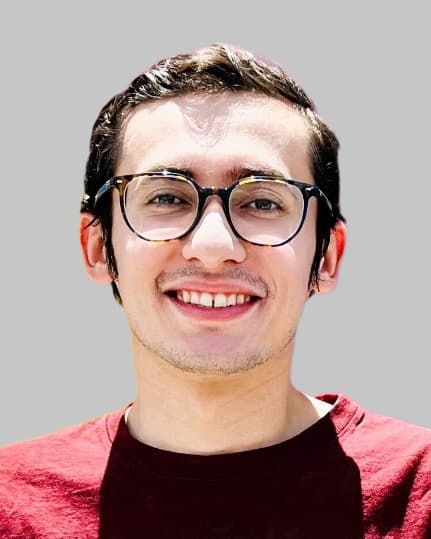}}]{Amirmohammad Mohammadi} received his M.Sc. in Electrical Engineering from Sharif University of Technology in 2021. He is currently pursuing a Ph.D. in Computer Engineering at Texas A\&M University. His research focuses on developing efficient deep learning architectures, specifically through parameter-efficient fine-tuning. Additionally, his work bridges fundamental artificial intelligence and digital health applications utilizing foundation models and physics-informed neural networks.
\end{IEEEbiography}

\begin{IEEEbiography}[{\includegraphics[width=1in,height=1.25in,clip,keepaspectratio]{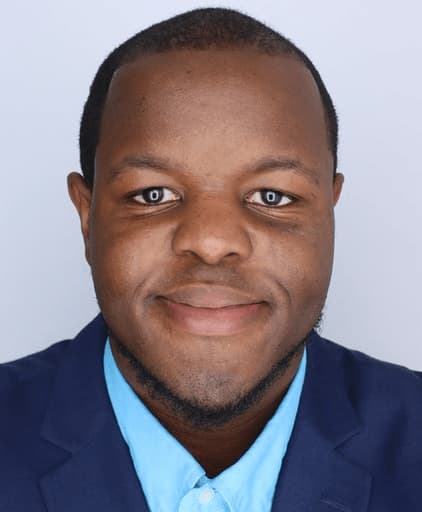}}]{Joshua Peeples} (Member, IEEE)
received the Ph.D. degree in electrical and computer engineering from the University of Florida in 2022. He is an Assistant Professor in the Department of Electrical and Computer Engineering at Texas A\&M University. His primary research interests include machine learning, computer vision, and image processing with a focus on image texture analysis.
\end{IEEEbiography}

\begin{IEEEbiography}[{\includegraphics[width=1in,height=1.25in,clip,keepaspectratio]{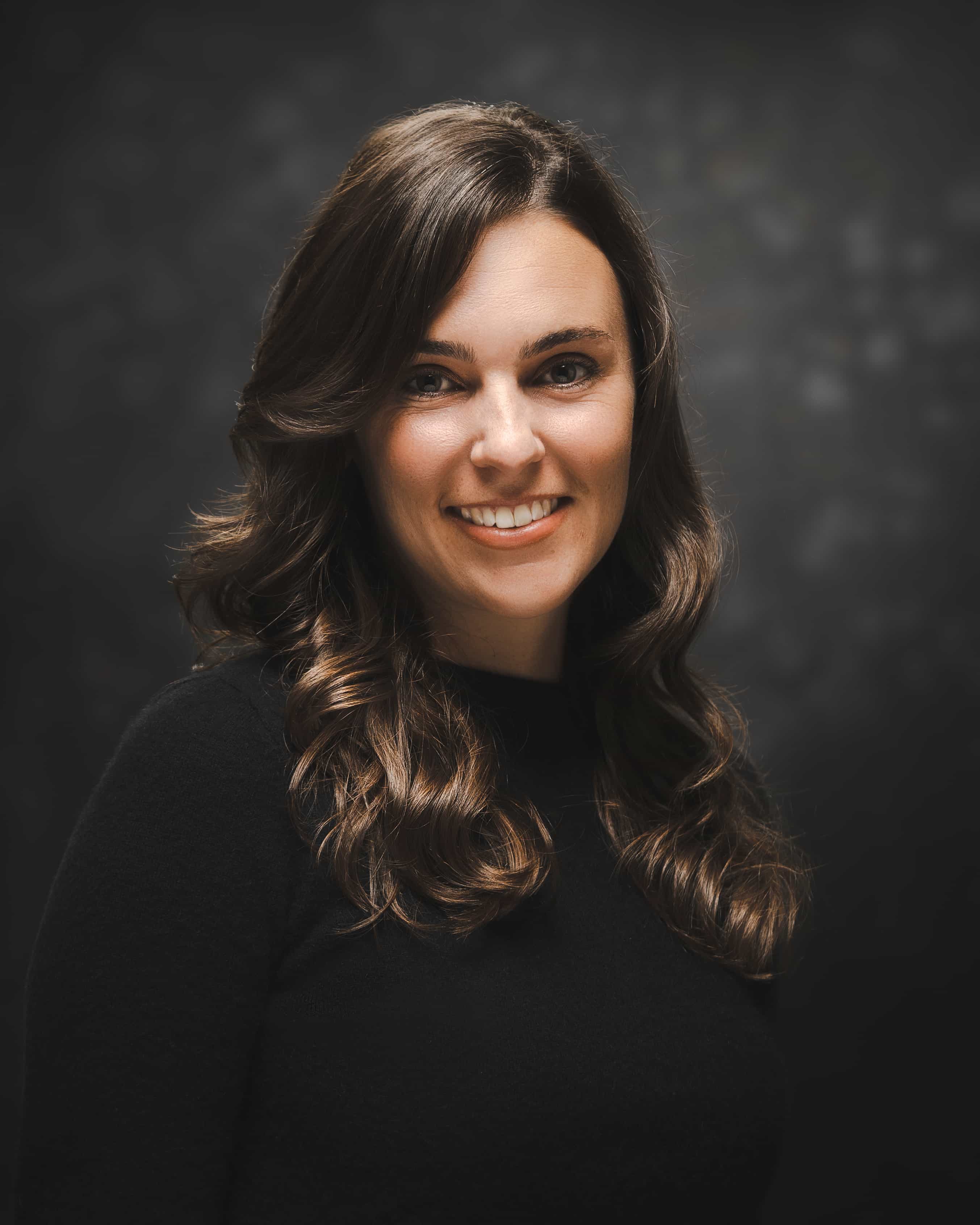}}]{Alexandra Van Dine} is an assistant group leader in the Advanced Undersea Systems and Technology group at MIT Lincoln Laboratory where she has interest in applied artificial intelligence and machine learning techniques in the development of maritime and undersea algorithms pertaining to acoustic signature detection and classification, anomaly detection, oceanographic forecasting, and object identification. Alexandra holds MS and PhD degrees in mechanical engineering from the University of California San Diego as well as a BS from the University of Virginia in aerospace engineering.
\end{IEEEbiography}

\vfill

\end{document}